\documentstyle[preprint,eqsecnum,aps,prb]{revtex}
\begin{document}
\draft
\preprint{}
\title{Theory of commensurable magnetic structures in holmium}
\author{Jens Jensen}
\address{\O rsted Laboratory, Niels Bohr Institute, 
Universitetsparken 5,
2100 Copenhagen, Denmark}
\date{\today}
\maketitle
\begin{abstract}The tendency for the period of the helically ordered
moments in holmium to lock into values which are commensurable with
the lattice is studied theoretically as a function of temperature and
magnetic field. The commensurable effects are derived in the
mean-field approximation from numerical calculations of the free
energy of various commensurable structures, and the results are
compared with the extensive experimental evidence collected during the
last ten years on the magnetic structures in holmium. In general the 
stability of the different commensurable structures is found to be in
accord with the experiments, except for the $\tau=5/18$ structure
observed a few degrees below $T_{\text N}$ in a $b$-axis field. The
trigonal coupling recently detected in holmium is found to be the
interaction required to explain the increased stability of the 
$\tau=1/5$ structure around 42 K, and of the $\tau=1/4$ structure
around 96 K, when a field is applied along the $c$-axis.

\end{abstract}

\section{Introduction}
\label{sec:level1}
The basic features of the magnetically ordered structures in hexagonal
close-packed Ho were established by Koehler {\it et al.},\cite{Koe1}\
who found that the basal-plane moments are arranged in a helical
pattern at all temperatures below $T_{\text N}\simeq133$ K. The
$c$-axis components order ferromagnetically at $T_{\text C}\simeq 20$
K, resulting in a conical ordering of the moments.
The opening angle of the cone was found to approach $80^\circ$ in
the zero temperature limit. The ordering wave vector for the
basal-plane moments is directed along the $c$-axis and its magnitude
$\tau$ (in units of $2\pi/c$) was observed to change 
monotonically from about 0.28 at $T_{\text N}$ to about $0.167\simeq1/6$ 
at $T_{\text C}$. Below $T_{\text C}$ the wave vector stays constant, 
indicating that the magnetic structure is locked to the lattice
periodicity and repeats itself after each twelve hexagonal layers.
The hexagonal anisotropy is known\cite{Yos} to produce higher harmonics at 
$(6\pm1)\tau$, and the fifth and seventh harmonics were
clearly resolved by Koehler {\it et al.}\ at low temperatures.
They concluded that the basal-plane moments in the cone
phase bunch strongly around the easy $b$-directions, so that the angle
between the basal-plane component of one of the moments and the
nearest $b$-axis is only about 5.8$^\circ$ in the zero temperature limit.
The neutron diffraction experiments have been repeated with higher
resolution and with crystals of better quality by Felcher
{\it et al.}\cite{Felch}\ and by Pechan and Stassis,\cite{Pechan}
leading to only minor modifications of the results of Koehler {\it et al.} 

Koehler and collaborators\cite{Koe2} also investigated the behavior
of the magnetic structure in holmium as a function of magnetic field
applied in the basal plane. The theoretical prediction for the
magnetization process was that the helix would change into a
ferromagnet either directly or via an intermediate fan
phase.\cite{Naga} This was confirmed to occur at low temperatures,
however, above 40--50 K Koehler {\it et al.}\ observed a second
fan-like phase.\cite{Koe2} A few years ago this extra intermediate
phase was explained, as being the helifan(3/2)-phase,\cite{J2,Jehan} where
the helifan structures are constructed from portions of the helix and
the fan following each other in a periodic way.     
        
The next major advancement in the investigation of the magnetic
ordering in holmium came with the use of the new technique of magnetic
x-ray scattering, which utilizes the intense radiation from a synchrotron
source. The narrow experimental resolution obtainable with this
technique made it possible for Gibbs {\it et al.}\cite{Gibbs1}\ to detect 
two other commensurable structures in holmium, with $\tau=2/11$ and 
5/27, in addition to the one with $\tau=1/6$, at temperatures below 25 K.
These commensurable values were explained by the spin-slip 
model.\cite{Gibbs1,Bohr} In the limit of very strong hexagonal
anisotropy the moments would be confined to be aligned along  
one of the six (in the present case) $b$-directions. 
The 12-layered structure may be constructed from pairs of neighboring
layers, in which the moments are along the same $b$-direction, 
by allowing the moments to rotate 60$^\circ$ from one pair to the next.
This periodicity may then be changed by introducing regularly-spaced 
series of {\it spin slips}, where at each spin slip a pair is replaced
by a single layer. For example, the 22 layered 2/11-structure may be
obtained from the 12-layered structure by introducing a spin slip 
after every 5 pair of layers, which we shall denote as the 
(222221)-structure. The introduction of a spin slip after every 4
pairs leads to the (22221)-structure corresponding to $\tau=5/27$.  
In this structure there is a spin slip for every 9 layers, and as
observed by Gibbs {\it et al.}, this gives rise to a modulation of the
lattice with this period, i.e.\ charge scattering at $\tau_c=2/9$.

The magnetic x-ray scattering technique provides very good resolution,
but the intensity is weak, thus preventing the measurement of the
scattering due to the higher harmonics. In contrast, neutron
diffraction reflections have high intensities, although the resolution
is relatively coarse. By the use of a large Ho-crystal, and a triple-axis 
neutron-scattering spectrometer for isolating the purely elastic
scattered neutrons, Cowley and Bates were able to determine
the intensities due to all the harmonics in a number of the spin-slip
structures, in which the intensities extend over four orders of
magnitude.\cite{Cow1} These results allowed them to derive the average
turn angle between the bunched pairs to be about 10$^\circ$ at the low
temperatures, in accordance with the result of Koehler {\it et
al.},\cite{Koe1} and this value increases to about 20$^\circ$ at 30 K.
In addition they were able to detect some variation of the turn angle
from pair to pair. The magnitudes of the two modifications of the
ideal spin-slip structure which were derived from the experiments were
reproduced in a numerical model calculation.\cite{Mac1} The ultrasonic
experiments made by Bates, {\it et al.}\cite{Bates} as a continuation
of the experiments of Cowley and Bates, showed anomalies in the sound
velocities not only at the temperatures where the system jumps between
the three commensurable structures described above, but also at about
25, 40 and 97 K. At these latter temperatures $\tau$ is close to the
commensurable values 4/21, 1/5, and 1/4, corresponding to the spin
slip structures (2221), (221), and (211), respectively.   

The possibility that the magnetic structure may lock into commensurable
values at elevated temperatures was investigated at Chalk River by
Tindall, Steinitz and collaborators.%
\cite{Stein1,Willis,Noakes,Tin1,Stein2,Tin2,Tin3,Tin4,Tin5,Tin7,Tin6}
They monitored the position of the fundamental magnetic diffraction
peak as a function of temperature at fields applied along the
$c$-direction or along the $b$-direction, and they found several
plateaus in the temperature variation of $\tau$. In the presence of 
a $c$-axis field of 30 kOe the (221)- and the (211)-structures were found
to be stable around 42 K and 96 K, respectively, in both cases
within a temperature interval of 2--3 K. In a $b$-axis field of 
14  or 30 kOe the (211)-structure is again stable for a couple of
degrees, which was also observed to be the case for the
(21)-structure, $\tau=2/9$, at about 75 K. Finally, they observed
$\tau$ to stay close to the value 5/18 between 126 K and 
the N\'eel temperature, when applying a field of 30 kOe along the $b$-axis. 
The observations made close to $T_{\text N}^{}$ and near 100 K have largely
been confirmed by the similar experiments of Venter and du 
Plessis,\cite{Venter1,Venter2} except that they did not observe a
clear plateau at $\tau=1/4$ in a $c$-axis field  around 96 K. However,
they detected an anomalous behavior of the scattering intensity close
to this temperature. 

The low temperature domain, below 40 K, has
been carefully investigated by Cowley {\it et al.}\cite{Cow2} in a 
$c$-axis field of 10--50 kOe. In spite of a rather monotonic
variation, with field or temperature, of the position of the
fundamental magnetic satellite they found by measuring the position of
the higher harmonics that the diffraction pattern was determined in
many cases by a superposition of neutrons scattered from domains with
different commensurable $\tau$-values. Small differences in $\tau$,
which may be difficult to resolve at the fundamental wave vectors, are
enhanced so to be distinguishable when considering e.g.\ the fifth or
seventh harmonics.

There have been made a number of theoretical studies of commensurable
structures in model systems. The simplest one, the anisotropic 
next-nearest-neighboring Ising (ANNNI) model has been investigated 
in detail both at zero field\cite{Fish,Bak1,Bak2} and in an applied 
field.\cite{Smith} The XY-model with six-fold anisotropy, 
which is more closely related to holmium, has recently been
discussed by Sasaki,\cite{Sas} in the limit of large anisotropy, and by
Seno {\it et al.}\cite{Seno} Steinitz {\it et al.}\cite{Stein3} have
made some qualitative considerations on the connection between the
spin-slip model and the field-dependent commensurable effects observed
in holmium. Plumer has analyzed a model, which includes most of the
couplings present in holmium, in order to understand the stability of
the 1/4-phase observed near 96 K in a $c$-axis field.\cite{Plumer} 
However, the explanation he proposed, relies on adding a large
symmetry-breaking term to the free energy which seems difficult to
justify. Here we shall present a mean-field calculation of the
stability of the different commensurable phases based on a realistic
model for holmium. The model includes not only the terms considered by
Plumer, but also the trigonal coupling which has recently been
detected in neutron diffraction measurements by Simpson {\it et
  al.}\cite{Simp} The trigonal coupling induces a net ferromagnetic
moment in the basal plane when $\tau=1/4$, and a misalignment of the
field by as little as one degree is sufficient for explaining the
observed lock-in of this structure, in accordance with the conjecture
made by Jensen and Mackintosh.\cite{J3} The basal-plane three-fold
anisotropy caused by the trigonal coupling in the case of a cone
structure has also strong implications on the (221)-structure in a
$c$-axis field near 42 K. 

The model used in the calculations is developed in the next section.
The stability of the different commensurable structures is
investigated in Sec.\ III. The free energy of the various structures
is calculated numerically within the mean-field approximation. The
advantage of this method is that it is possible to account for most of
the complexities of the real system, whereas one disadvantage is that
there is a limit to the number of different layers one may handle 
numerically with the sufficient precision. In the present calculations 
we consider structures with a repeat length of up to 500 layers,
leading to a resolution which should be superior to even the most
precise experiments. The work is summarized and concluded in the last
Sec.\ IV.

\section{The mean-field model of holmium}
\label{sec:level2}

The model is based on the magnetization and spin-wave measurements, 
and is similar to the one applied in previous numerical
analyses of the magnetic structures and excitations in 
holmium.\cite{J2,Mac1,J3,Lars,J1,J4,Mac2} The Hamiltonian comprises 
the single-ion anisotropy, the Zeeman term, the Heisenberg exchange
coupling, the classical magnetic dipole--dipole interaction, and the 
trigonal coupling:
\begin{eqnarray}
{\cal H}_0^{}=&&\sum_i\sum_{lm}B_l^mO_l^m(i)
-\sum_ig\mu_B^{}\text{\bf H}\cdot\text{\bf J}_i\nonumber\\
&&{}-\textstyle{1\over2}\sum_{ij}
{\cal J}(ij)\,\text{\bf J}_i\cdot\text{\bf J}_j
-{\textstyle{1\over2}}\sum_{ij}
{\cal J}_{\!D}^{}(ij)J_{zi}J_{zj}\nonumber\\
&&{}+\sum_{ij}K{}_{21}^{31}(ij)
\big[O_3^2(i)J_{yj}+O_3^{-2}(i)J_{xj}\big].
\label{1}
\end{eqnarray}
The $O_l^m$-operators are the Stevens operators, and in particular 
$O_3^{\pm2}={1\over2}(J_zO_2^{\pm2} +
O_2^{\pm2}J_z)$, where $O_2^2=J_x^2-J_y^2$ and $O_2^{-2}=J_x^{}J_y^{} 
+J_y^{}J_x^{}$. The $x$-, $y$-, and $z$-axes are assumed to be
along the $a$-, $b$-, and $c$-axes of the HCP lattice, respectively.
The Fourier transforms of the couplings are defined in the standard
way, and we shall use the short hand notation ${\cal J}(q)$ for 
${\cal J}(\text{\bf q})$, when {\bf q} is along the $c$-axis. The
inter-planar exchange parameters ${\cal J}_n$ are then defined by
\begin{equation}
{\cal J}(q)={\cal J}_0+2\sum_{n \ge 1}{\cal J}_n\cos(nqc/2).
\label{2}
\end{equation}
The classical dipole interaction is included in the coupling ${\cal J}(q)$
of the basal-plane moments, in which case the coupling of the
$c$-components is ${\cal J}_{cc}(q)={\cal J}(q)+{\cal J}_D(q)$. 
The classical contribution vanishes at zero wave vector,
${\cal J}_D(\text{\bf 0})\equiv 0$, whereas at $q\ne0$
\begin{eqnarray}
&&{\cal J}_D(q)=\nonumber\\
&&{}-{\cal J}_{dd}\{0.919+0.0816\cos(qc/2)-0.0006\cos(qc)\},
\label{3}
\end{eqnarray}
where the coupling constant in holmium is
\begin{equation}
{\cal J}_{dd}=4\pi(g\mu_B^{})^2\,N/V= 0.0349\hbox{\ meV.}
\label{4}
\end{equation}
The jump, which the dipole coupling ${\cal J}_D(q)$ makes at zero wave
vector, is observable in the excitation spectrum, and it explains
without introducing any further two-ion anisotropy,\cite{Lars} why the
system prefers the cone structure rather than the ``tilted helix''
below $T_{\text C}^{}$.

The trigonal coupling was discovered to be present in erbium by Cowley
and Jensen.\cite{Cow3} This coupling reflects the fact that the
$c$-axis is a three-fold symmetry axis, and its contribution to the
mean-field Hamiltonian for the $i$th ion is given by 
\begin{eqnarray}
&&\Delta{\cal H}_{\text MF}^{}(i\in p\hbox{'th plane})=
(-1)^p\sum_{n\ge1}\big[K{}_{31}^{21}\big]_n^{}\nonumber\\
&&{}\times\Big[\{O_3^2(i)-\textstyle{1\over2}\langle O_3^2(i)\rangle\}
\big\langle J_y^{}(p+n)-J_y^{}(p-n)\big\rangle\nonumber\\
&&{}+\{O_3^{-2}(i)-\textstyle{1\over2}\langle O_3^{-2}(i)\rangle\}\big\langle 
J_x^{}(p+n)-J_x^{}(p-n)\big\rangle\nonumber\\
&&{}-(-1)^n\{J_{yi}^{}-\textstyle{1\over2}\langle J_{yi}^{}\rangle\}
\big\langle O_3^{2}(p+n)-O_3^{2}(p-n)\big\rangle\nonumber\\
&&{}-(-1)^n\{J_{xi}^{}-\textstyle{1\over2}\langle J_{xi}^{}\rangle\}
\big\langle O_3^{-2}(p+n)-O_3^{-2}(p-n)\big\rangle\Big],\nonumber\\
\label{5}
\end{eqnarray}
where the argument $p\pm n$ denotes an ion in the uniformly magnetized
$(p\pm n)$th hexagonal layer. This equation, which defines the
inter-planar coupling parameters $\big[K{}_{31}^{21}\big]_n^{}$, shows
that the coupling changes sign from one sublattice to the next. Both
in the cone phase and in the cycloidal phase of erbium, the trigonal
coupling gives rise to additional neutron diffraction peaks when the
scattering vector is along the $c$-axis. These peaks would not be there
if the magnetic structures were independent of the different
orientation of the basal planes in the two hexagonal sublattices. 
The equivalent phenomenon has also now been observed in holmium by
Simpson {\it et al.},\cite{Simp} who fitted the intensities of the 
extra reflections by a set of three inter-planar parameters for the
trigonal coupling.

We have reanalyzed the experiments of Simpson {\it et al.}\cite{Simp}
and derived a set of anisotropy parameters, which accounts 
both for the low temperature magnetization curves, as precisely as the
previous model,\cite{Lars} and for the neutron diffraction results. 
The effects of all the three trigonal couplings of the fourth 
rank,\cite{Cow3} and combinations of the three couplings
were investigated, and in agreement with Simpson {\it et al.}\ we
find that the trigonal coupling introduced by Eq.\ (\ref{5}) (the 
coupling applied in erbium), is the one which leads to the best fit.
The small differences between the ways the neutron diffraction 
results are fitted are insignificant compared with the experimental
uncertainties, but the inter-planar coupling parameters derived here
are, quite remarkably, a factor of 4-5 smaller than those obtained by 
Simpson {\it et al.} This large difference is surprising, but may be
explained by the high degree of compensation which occurs between the
different terms in the mean-field Hamiltonian (\ref{5}). The degree of
compensation depends on the structure considered, and in the analysis
of the commensurable structures discussed in the next section we
found that the trigonal coupling derived by Simpson {\it et al.}\ 
has unacceptably strong implications in some cases. The alternative
trigonal coupling derived here circumvents these difficulties, and is
therefore a more likely possibility. The modified set of 
inter-planar coupling parameters for the trigonal coupling is given in
Table I. The corresponding crystal-field parameters $B_l^0$ are
derived from the low-temperature magnetization curves and $T_{\text
  C}$, and $B_6^6$ is determined so that the averaged bunching angle
is $5.8^\circ$ in the 12-layered structure in the low temperature
limit. The values of these parameters are given in Table II, and they
are close to those applied in the previous model.\cite{Lars}
 
The variation of the ordering vector with temperature in Ho has been
analyzed by Pechan and Stassis,\cite{Pechan} who found that it agrees
reasonably well with the prediction of the theory of Elliott and
Wedgwood.\cite{Elliott} ${\cal J}(\text{\bf q})$ is proportional to 
the susceptibility of the conduction electrons, which depends
strongly on the nesting between different parts of the Fermi surface.
The super-zone energy gaps created due to the oscillating polarization
of the conduction electrons lead to a decrease of the maximum in the
susceptibility and to a shift of its position towards smaller wave
vectors, as the degree of polarization is increased. In addition, 
Andrianov\cite{And} has discovered a relation,
$\tau=\tau_0^{}[(c/a)_{\text cr}- c/a]^{1/2}$, between the $c/a$-ratio
and the ordering wave vector, where $(c/a)_{\text{cr}}\simeq 1.582$.
This relation is obeyed nearly universally by the 
rare-earth metals, indicating a strong correlation between the
$c/a$-ratio and the nesting effects. A change of the $c/a$-ratio
will modify the magnetic Hamiltonian in ways other than through 
${\cal J}(\text{\bf q})$. $B_2^0$ is especially sensitive 
to the $c/a$-ratio, as to a first approximation it is proportional to
the difference between $c/a$ and its ideal value, and rather strong 
effects have recently been discovered in thulium, when the system jumps
from the ferrimagnetic phase with $\tau=2/7$ to the ferromagnetic
structure.\cite{Keith} In the model considered here we only account
for the temperature variation in ${\cal J}(q)$, whereas additional
magneto\-elastic effects on the crystal-field parameters or changes of
${\cal J}(q)$ as a function of field are neglected. The spin-wave
dispersion relation of Ho has been measured both at low
temperatures\cite{Lars,String,Patt} and at elevated
temperatures,\cite{Nick} and these measurements have been used to derive
the inter-planar exchange parameters.\cite{Lars,J1} In order to reproduce
the correct wave vectors at the different temperatures small adjustments
have been introduced, and for this purpose we have
chosen ${\cal J}_3$ as the variation parameter. The inter-planar
exchange parameters used in the model are given in Table III.

The structures discussed in the next section have been calculated by a
straightforward iteration procedure using the parameters given in
Table I-III and Eqs.\ (\ref{3})-(\ref{4}). The calculations utilize
the mean-field approximation, in which $J_{\alpha i}J_{\alpha j}$ in
(\ref{1}) is replaced by $(2J_{\alpha i}-\langle J_{\alpha
  i}\rangle)\langle J_{\alpha j}\rangle$ and the trigonal term by Eq.\
(\ref{5}). The first step in the iteration is to assume a 
distribution of expectation values of the various operators. These
values are then inserted in the mean-field Hamiltonian for the $i$th ion,
and after a diagonalization of this Hamiltonian the partition function, the
free energy, and new expectation values for this ion are calculated.  
The calculation is carried out for each ion in one
commensurable period, and the procedure is repeated with the new 
distribution of expectation values until a self-consistent solution is
achieved. At a given temperature and field the free energies of
structures with different (commensurable) periods are
compared in order to identify the most stable structure. The energy
differences between the various structures may be minute, and
the calculations have to be done with high numerical accuracy. In
order to ensure in a given case, that the iteration has converged 
towards the state with the lowest free energy and not to a metastable 
configuration, many iterations are required, of the order of several 
thousands, and many starting configurations have to be considered.

\section{The commensurable structures}
\label{sec:level3}
\subsection{Low temperature regime}
Most of the commensurable structures are observed in the low
temperature regime ranging from zero to about 50 K. The exchange
constants are given in Table III at the end points of this
interval. In between we have used the quadratic interpolation:
\begin{equation}
J_n(T)=(1-\alpha^2)J_n(0)+\alpha^2 J_n(50 \text{\ K}), 
\label{6}
\end{equation}
where $\alpha$ is the temperature $T$, in units of K, divided by 50. 

At zero field the lowest order spin-slip structures are particularly
stable. Starting at about 40 K, we get the sequence (221), (2221),
(22221), (222221), corresponding to $\tau=1/5$, 4/21, 5/27, and 2/11,
with a steady increase of the  number of pairs in between the
spin-slip layers as the temperature is reduced to about 20 K. One
might expect this series to continue as the system is cooled further,
but the 2/11-structure is predicted to occur only in a small
temperature interval of about 0.8 K, and the intervals, where the
subsequent spin-slip structures are stable, become minute. Thus the
system is predicted to jump from the 2/11-structure to the pure  
pair-layered  1/6-structure within an interval of 0.1 K. 
Metastable states, or mixed states are expected to appear frequently
in the real system, which makes it difficult to decide whether this  
particular prediction is in accord with the experiments, but the
temperature interval, in which the indications\cite{Cow1,Bates} of
the intermediate structures are found is narrow with a width of about
1 K. 

In addition to the lowest order spin-flip structures, structures with
mixed sequences are also found to be stable. With one exception, the 
(222122221)-structure ($\tau= 3/16$), which is stable over a small
interval of about 0.3 K, all the higher order spin-slip structures 
occur between the (221)- and the (2221)-structures, of which the most
important one is the (2212221)-structure ($\tau=7/36$). 
When approaching $\tau=1/5$ from below, the main sequence of
commensurable ordering vectors is determined by $\tau_n(1)=n/(5n+1)$
which is the series of rational fractions close to 1/5 with the
smallest denominators or shortest commensurable periods. Commensurable
structures with wave vectors lying in between those determined by
$\tau_n(1)$ may also be stable. In order to investigate this in a
systematic way we have used the proposal of Selke and
Duxbury that the structure most likely to appear between two phases
has a period which is the sum of the periods of the structures in the
two phases,\cite{Seno,Selke} suggesting that the 
most probable intermediate values in the present case are those
determined by the second-order series $\tau_n(2)=(2n+1)/(10n+7)$. For
a further subdivision of the steps made by $\tau$ we have applied the 
third-order series
$\tau_n(3)=(3n+1)/(15n+8)$. In the interval between $\tau=4/21$ 
and $\tau=7/36$ the most stable configuration is found to  
be the $\tau_5^{}(2)=11/57$ or (22122212221)-structure, whereas other
choices from the $\tau_n(1)$ or the $\tau_n(2)$ series are only found
to be stable in narrow temperature intervals. For $\tau$ larger than
7/36, the situation is changed. Structures defined by the series
$\tau_n(1)$ as well as $\tau_n(2)$ are stable in about
equal intervals, and for $n\ge9$, structures determined by $\tau_n(3)$
also appear. At these temperatures the intervals are so small that it
is difficult by the numerical calculations to distinguish the
behavior from a truly continuous, incommensurable, variation of
$\tau$. The magnetic correlation length in the $c$-direction of Ho has
been determined by x-ray scattering\cite{helg} to be of the order of
2000 layers, indicating that if the shortest commensurable period
becomes of the order of 200 layers, it is no longer possible
experimentally to distinguish commensurable ordering, repeating itself
coherently about ten times, from an incommensurable phase. The
non-zero resolution characterizing the numerical calculations is a
limitation, when comparing the results with more exact analytic
results, but not in a comparison with a real system, which suffers
from impurities, stacking faults and other imperfections. Hence we
expect that experimentally $\tau$ will change continuously, at least
when it is larger than about 0.197 until it, at a first order
transition, jumps to the commensurable value $\tau=1/5$. The model
calculations predict a lock-in at this wave vector between 41.47 K and
43.70 K, and that $\tau$ returns to incommensurable values above 43.70
K. The calculated results for the variation of $\tau$ at zero field,
in the low temperature regime, are shown in Fig.\ 1. 

The stability of the different commensurable structures are influenced
by a field along the $c$-axis. The calculations indicate that the main
features of the phase diagram are the same as at zero field, but there
are a few significant changes. There is an overall shift of the
stable regimes of the commensurable structures towards lower
temperatures, the more pronounced the lower the temperature is.
The 2/11-phase and the spin-slip structures lying in between this
phase and the 1/6-phase become more stable, until the field is so
large that the overall temperature shift removes these phases, which
happens between 25 and 35 kOe. Finally, the most important change is,
that the stability of the $\tau=1/5$-phase is much increased by 
the field. 

The first of these effects is explained by the condition that the
moments in the spin-slip layers have a larger $c$-axis susceptibility
than the moments in the pair layers, i.e.\ the more spin-slip layers a
structure contains the more Zeeman energy it gains in a $c$-axis
field. The experimental results of Cowley {\it et al.},\cite{Cow2}
obtained at a $c$-axis field of 10--50 kOe, which are included in Fig.\ 1,
reflect this effect. Fig.\ 2 shows a more direct comparison of the
experiments of Cowley {\it et al.}\ with the calculated field
dependence of the ordering wave vector at 10 K. The experimental
results shown in Fig.\ 2 are derived from the position of the primary
magnetic satellite near $(002)$, which leads to a more gradual change of
the wave vector than if the contributions from different domains with
different stable or metastable structures are separated, as shown in 
Fig.\ 1. The comparison in Fig.\ 2 demonstrates that the calculated shift of
the ordering wave vector at a given temperature (i.e.\ at fixed values
of the coupling constants) due to the $c$-axis field is of the same
magnitude as observed experimentally. The calculated results at zero
temperature are similar to the results in Fig.\ 2 at 10 K, except
that the values of the transition fields are shifted upwards by about 
5 kOe. The strong field-dependence of $\tau$ should be taken into
consideration in the comparison between the experiments and the
theoretical results in Fig.\ 1. In this comparison it may also be
important that the 2/11-structure, in contrast to most of the other
spin-slip structures, has a moment in the basal plane, and that
the 3/16-structure, in between the 5/27- and the 4/21-structures, is
much more susceptible to a basal-plane field than the two neighboring
structures. This means that a small misalignment of the $c$-axis field
would favor principally these two structures.

At not too high temperatures, the hexagonal anisotropy energy decreases
with the relative magnetization, $\sigma$, proportional to
approximately $\sigma_{}^{21}$, whereas the trigonal anisotropy
changes like $\sigma_{}^7$. The hexagonal anisotropy dominates at the
lowest temperatures, and is responsible for the commensurable 
spin-slip structures, but as the temperature is increased the trigonal 
anisotropy becomes relatively more important. At 40 K 
$\sigma\simeq0.925$, implying that the hexagonal anisotropy energy has
decreased by a factor of 5 compared with its value in the zero
temperature limit, whereas the trigonal anisotropy is only reduced by
a factor of 1.7. Within perturbation theory the 
contribution of the trigonal interaction, Eq.\ (\ref{5}), to the free
energy is of second order in the helical phase, whereas a first-order
contribution appears if the $c$-axis moments are nonzero. To a
first approximation this contribution is proportional to
\begin{equation}
\Delta F\propto \sum_p(-1)_{}^pJ_\parallel^{}J_\perp^3\cos(3\phi_p),
\label{7}
\end{equation} 
where $J_\parallel^{}$ and $J_\perp$ are the components of the moments
parallel and perpendicular to the $c$-axis, respectively, and $\phi_p$
is the angle the perpendicular component of the moments in the $p$th
layer makes with the $x$- or $a$-axis. When only the trigonal coupling
is considered, then every second $a$-axis is an easy axis in one of
the sublattices and the other three $a$-axes are the easy axes in the
other sublattice. Fig.\ 3 shows the hodographs of the basal-plane
moments calculated at 42.185 K in zero field and in the presence of a
field of 10 kOe along the $c$-axis. The figure indicates that the
three-fold anisotropy term induced by the $c$-axis field, (\ref{7}), 
is capable of
rotating the moments about 30$^\circ$, so that the moments in the two
spin-slip layers, which at zero field are along a $b$-axis become
oriented along an $a$-axis. The hexagonal anisotropy energy does not
depend on the distinction between the two sublattices and is changed
by the same amount as obtained by rotating the zero-field hodograph the
relatively small angle in the opposite direction. The small rotation
does not cost much in hexagonal anisotropy energy, whereas the gain in
trigonal anisotropy energy is substantial. 

Both structures in Fig.\ 3 look like spin-slip structures, except that
the roles of the $a$- and the $b$-axes have been interchanged in the
two cases. However, at these temperatures the basal-plane anisotropy
energy is reduced so much that the spin-slip model is no longer
particularly useful, since the angle between the moments in the pair
layers is not much smaller than the smallest angle between moments
belonging to neighboring pairs. The commensurable structures with
$\tau$ slightly different from 1/5 are not spin-slip structures.
Instead they consist of portions of the 1/5-structure separated by
domain walls in which phase shifts are 
introduced via a relatively smooth adjustment of the turn angles. 
If the width of the walls is much smaller than the distance between
them, then the coupling between the walls is negligible and the free
energy changes linearly with the density of walls,
corresponding to a linear variation of the free energy with $\tau$. A
detailed discussion of the behavior of systems with domain walls is
given by Fisher and Szpilka.\cite{Fisher} Fig.\ 4 shows the free
energy calculated in the two cases, for various commensurable values
of $\tau$ in the proximity of $\tau=1/5$. In the case of
$H_c=10$ kOe, the free energy varies linearly between 0.198 and 0.2 and
between 0.2 to 0.202, whereas the variation is parabolic at zero field,
except that the free energy  at $\tau=1/5$ is slightly smaller than
the minimum value of the parabola. Leaving out the result at $\tau=1/5$,
the free energies of the zero-field commensurable structures are fitted
by a third degree polynomial with a standard deviation of about
$10^{-7}$ meV, which is the curve shown in the figure, (the numerical
accuracy by which the free energy is calculated is $10^{-8}$ - $10^{-9}$
meV). This smooth change of the free energy
is consistent with the estimated incommensurable variation of $\tau$ on 
both sides of the temperature interval where the $\tau=1/5$-structure 
is stable. The free energy in the zero field case is nevertheless
going to change linearly with $\tau$ for $\tau$ sufficiently
close to 1/5, when $|\tau-1/5|$ is smaller than about 0.0007. 

The difference between the behavior of the free energy at zero and
at $H_c=10$ kOe is related to the different widths of the domain walls. 
Fig.\ 5 shows the turn angles in the two cases calculated at 42.815 K.
In both cases $\tau$ is slightly smaller than 1/5 corresponding to an
average turn angle slightly smaller than 36$^\circ$. The width of the
domain walls, denoted by the arrows, is about 100 layers in zero field
and about 70 layers at $H_c=10$ kOe. However, the phase shift
accomplished by one domain wall is different in the two cases:
$12^\circ$ in the zero-field case and 24$^\circ$, a factor 2
larger, in the case of $H_c=10$ kOe. These numbers imply that 
the domain walls at zero field start to overlap with each other when 
$1/5-\tau$ is larger than 0.0007, whereas when $H_c=10$ kOe the overlap
starts to occur only when $1/5-\tau$ is nearly a factor of 3 larger.  
These same numbers also describe the situation at $\tau$-values larger
than 1/5.

At decreasing temperatures the slope of the linear section in
the free energy decreases, numerically, for $\tau$ smaller than 1/5,
and the opposite for $\tau$ larger than 1/5. The 1/5-phase is the
stable one until the temperature is lowered to the point
where the slope becomes zero. At this temperature it costs no energy 
to create domain walls, as long as the distance between them is larger 
than their width. This leads to a vertical decrease of the 
$\tau$-value at this temperature. At a slightly lower temperature the 
density of the domain walls will be so large that the walls
overlap each other, and there will be only one equilibrium 
configuration, corresponding to the curvature in the free-energy
function at the minimum being non-zero. The change of $\tau$ as a
function of temperature will accordingly be more gradual. The
transition between the 1/5-structure and other structures is 
predicted to occur in a (quasi)continuous\cite{Fisher} way in both the
cases considered, but at zero field the transition regime is so
narrow that the transition is indistinguishable from a first-order
one, within the present numerical resolution.    

The lock-in of the structure at $\tau=1/5$ is strongly enhanced by the
$c$-axis field. As shown in Fig.\ 6, the interval in which the
1/5-structure is calculated to be stable increases rapidly, from about
2.2 K to about 10 K, between zero and a field of 10 kOe along the
$c$-axis. This strong increase is also perceptible on either side of
the 1/5-phase. Here commensurable effects are resolved even
very close to the transition regions at 10 kOe, in contrast to
the incommensurable behavior indicated by the calculations at zero-field.  
The large field-induced enhancement of the commensurable effects is
produced solely by the trigonal coupling. We have repeated the
calculations using the previous model\cite{Lars} which neglects the
trigonal coupling. In this case the zero-field structure is found to
be stable within an interval of 2.8 K, and this interval is reduced 
in a $c$-axis field, by a factor of 2 at a field of 30 kOe. The effective 
hexagonal anisotropy decreases faster than the free-energy differences
between the different structures when the basal-plane moments are
reduced by the field. 

Tindall {\it et al.}\cite{Tin5} have observed a
plateau in the variation of $\tau$ at 1/5 over a temperature range of
2--3 K when $H_c=30$ kOe, see Fig.\ 1, which disappears at
zero field. It is probably very difficult to obtain experimental
results on commensurable effects which can be trusted quantitatively
in a comparison with the results derived from a mathematical model of
an ideal system. The commensurable effects are based on very small
differences in the free energy and impurities and defects are likely to
be important. Furthermore, the results of Cowley {\it et
al.}\cite{Cow2} in Fig.\ 1 show that, at low temperatures, several
metastable configurations may be present in the crystal at the same
time. The occurrence of metastable structures should be most
pronounced at low temperatures, as also indicated by the behavior of
the hysteresis effects, however, mixed phases are still observed at 
35 K, see Fig.\ 1. Taking these effects into consideration, it seems
reasonable that the calculated lock-in at zero-field of the 
1/5-structure within an interval of about 2 K, may be smoothed out in
the temperature variation of the average position of the fundamental
magnetic scattering peak. We may instead refer to the low-field
anomalies observed near 40 K in ultrasonic\cite{Bates}
and magnetization\cite{Willis} measurements. The observed
field enhancement of the lock-in of the 1/5-phase is in contradiction
with the behavior expected if the trigonal coupling were negligible.
Zeeman effects due to a small misalignment of the $c$-axis field are
estimated to be of no importance, so, qualitatively, the enhancement
may only be explained by the trigonal coupling. The question left open 
is whether the observed plateau of about 2 K at 30 kOe is consistent
with the calculated range of the lock in of about 10 K.

\subsection{High temperature regime}
The high temperature regime is here considered to extend from 
40--50 K and up to $T_{\text N}$. In this temperature range the
anisotropy within the basal plane is relatively weak, and the
spin-slip model no longer applies. Instead the additional
anisotropy introduced by an external field will play a much more
important role. The value of $\tau$ changes continuously in most of
the interval, and detectable commensurable effects are only expected 
for structures with short commensurable periods, and we have limited
ourselves to the study of the following three cases; the 18-layered
2/9-structure, the 8-layered 1/4-structure, and the 36-layered 
5/18-structure appearing respectively around 75 K, 100 K, and 130 K.

Assuming the average slope $d\tau/dT$ near 75 K to be 0.001 K$^{-1}$,
we calculate the lock-in interval of the 2/9-structure to be 0.5 K at
zero field. This value is nearly unchanged in a field of 30 kOe along
the $c$-axis. According to the magnetization measurements the
transition to the fan structure, via intermediate helifan structures,
occurs at a field of about 18 kOe applied along a $b$-axis.\cite{Feron,Mac1} 
At 14 kOe the system is still in the helical phase, but the helix
is strongly distorted by the field, the magnitude of the second (or
seventh) harmonic of the basal-plane moments has grown to be 
almost the same as in the fan phase, and the 2/9-structure is found to
be stable in an interval of 3.6 K. These commensurable effects for
the 2/9-structure are found to be nearly independent of the trigonal
anisotropy. The only weak tendency for the system to lock in at
$\tau=2/9$ at zero field or in the presence of a $c$-axis field seems 
to be in accordance with experiments. Tindall and collaborators have
not reported any lock in of the 2/9 structures in these two cases,
whereas they observed a plateau in the $\tau$ variation with a width of 
about 2 K in the presence of a $b$-axis field of 14 kOe,\cite{Tin6}
which is consistent with the calculated behavior.

In the helical phase, the field perpendicular to the helical axis leads
to a slight reduction of the $\tau$-value. The transition to the
fan-phase is accompanied by a further reduction of $\tau$, hereafter
in the fan phase there is a  
steady increase of $\tau$ with increasing field until the system
becomes ferromagnetic. The possibility that the ordering wave vector
may change was not included in the original analysis of the helix-fan
transition by Nagamiya {\it et al.}\cite{Naga} It has been observed
by Koehler {\it et al.}\cite{Koe2} at 50 K, where $\tau$ changes
from 0.208 at zero field to be 0.170 in the fan phase, a change which
was found to agree with the model calculations.\cite{J2} At 80 K
$\tau$ is about 0.230 at zero field and is calculated to be reduced to
0.199 when the helical structure is changed into a fan at 18 kOe. In
the fan phase $\tau$ increases and is 0.222 at 30 kOe at the transition
to the ferromagnet. The magnitude of these changes in $\tau$ decreases
linearly with increasing temperatures and is a factor of 3 smaller at
100 K. This means that the temperature dependence of $\tau$ in the
interval between 80 and 100 K is a factor of 2 larger at a
field of 20 kOe, just above the helix-fan transition, than at zero field.  
This prediction may be compared with the experimental results of 
Tindall {\it et al.}\cite{Tin4} shown in Fig.\ 5 of their paper. This
figure shows that $\tau$ changes from about 0.18 at 81 K to about 0.23
at 95 K, when a field of 30 kOe is applied along a $b$-axis. This change
is a factor of 3 larger than indicated by the calculation at 30 kOe,
but it is comparable with the prediction at 20 kOe. In the fan phase the
average basal-plane moment is about 6 $\mu_{\text B}^{}$, which leads
to a demagnetization field in the case of a sphere of about 8 kOe, and
it is therefore quite likely that the applied field of 30 kOe should be
reduced by about 10 kOe in the comparison with the model calculations. 
Domain effects combined with the large value of the demagnetization
field may also be the reason why Tindall {\it et al.}\cite{Tin4}
observed the magnetic satellite to split into two peaks in some of the
scans, the observed splitting in $\tau$ of 0.007 corresponds to
a field difference of 4--5 kOe. Tindall {\it et al.}\ do not
report any particular lock-in effects for the structures shown in
their Fig.\ 5. This is consistent with the model calculations which
also do not indicate any significant commensurable effects for these
structures.  

The change of $\tau$ as a function of a basal-plane field is still of
some magnitude around 100 K, and the calculated effect agrees
well with the observation of Venter {\it et al.}\cite{Venter1} At 100
K the helix-fan transition is predicted to 
occur at a field of 18 kOe. At fields just below this value the
1/4-phase is found to be stable within an interval of about 2 K, 
the interval is 2.0 K at 16 kOe assuming $d\tau/dT=0.00128$ K$^{-1}$, 
whereas this interval is reduced by a factor of two or more in the fan
phase. Tindall {\it et al.}\cite{Tin2,Tin4,Tin6} have observed a
lock-in of the 1/4-structure in an interval of about 1 K at 14 kOe and
about 1.5 K at 30 kOe, and Venter {\it et al.}\cite{Venter1} have made
the similar observation that the 1/4-structure is stable within an
interval of about 1 K at $H_b=17$ and 23 kOe. The system is most
likely to be still in the helical phase at 14 and 17 kOe. The
positions of these lock-ins, just below 100 K in both cases, and their
widths are in good agreement with that predicted in the helical case.
The present model indicates that the stability of the 1/4-structure
should decrease when the system jumps into the fan-phase, which is
somewhat in disagreement with the experiments. We have two remarks to
add to this minor discrepancy. Firstly, the (effective) anisotropy
parameters are determined from the  behavior of the system at low
temperatures, and their values may have changed at these high
temperatures. An increase of the axial anisotropy $B_2^0$ stabilizes
the 1/4-structure in the fan phase, but does not have much effect on
the helical structure. Secondly, it is uncertain whether the system is
in the fan phase at the fields of 23 and 30 kOe. The relatively large
demagnetization field in the case of the fan may lead to a mixing of 
the phases, and in addition there is the likely possibility that 
the structures are helifans,\cite{J2} in which case the commensurable
effects are larger than they would be in the fan phase.                

Around 100 K the hexagonal anisotropy energy is very small and the
calculations indicate that the basal-plane turn angle at zero field
only differs by up to 0.12$^\circ$ from the average value of
45$^\circ$ for the 8-layered, $\tau=1/4$, structure at zero field. 
The application of a field along the $c$-axis implies that the
first-order trigonal contribution, Eq.\ (\ref{7}), becomes non-zero
and the trigonal anisotropy leads to a variation of the turn angle
between 44.5$^\circ$ and 45.5$^\circ$ at $H_c=30$ kOe. This
modification is still small, and the model calculations indicate only
very weak commensurable effects, a lock-in temperature interval of
the order of 0.1 K both at zero field and when $H_c=30$ kOe. In
analogy with the fifth and seventh harmonics induced by the hexagonal 
anisotropy,  the first-order trigonal anisotropy induces a second and
a fourth harmonic, but because of the factor $(-1)^p$ in Eq.\ (\ref{7})
these harmonics are translated by a reciprocal lattice vector along the
$c$-axis (the half of a reciprocal lattice vector in the double-zone
scheme), which means that the fourth harmonic appears at zero wave 
vector when $\tau=1/4$. In other words, in the case of a cone
structure with $\tau=1/4$ the trigonal coupling leads to a
ferromagnetic component perpendicular to the cone axis. The two extra
peaks observed in the cone phase in, for instance, the
2/11-structure\cite{Simp} may be classified as arising from these
harmonics, and in the cone phase of erbium, where $\tau=5/21$ is close
to 1/4, the fourth harmonic is observed close to the nuclear
peak,\cite{Lin} see also the discussion in Ref.\ [\onlinecite{Cow3}].
In the present case the ferromagnetic moment is along a $b$-axis and
is calculated to be 0.043 $\mu_{\text B}^{}$ when $H_c=30$ kOe,
corresponding to an average rotation of the moments towards the
$b$-axis by about 0.4$^\circ$. 

The estimated moment is small, but the energy differences determining
the commensurable effects are also small, and the basal-plane
moment has a profound influence on the system, as soon as the field
has a non-zero component perpendicular to the $c$-axis. The Zeeman
energy gained by the 1/4-structure, in comparison with the neighboring
structures, is proportional to the angle $\theta$ between the $c$-axis
and the direction of the field, at small values of the angle.
At the temperature where the $\tau=1/4$ structure is stable at
$\theta=0$, the free energy of the structures increases quadratically
with $\tau-1/4$, which leads to a lock-in temperature interval,
$\Delta T$, of the 1/4-structure proportional to $\sqrt{\theta}$. 
In this simple estimate we have neglected the small commensurable
effect at $\theta=0$, and the possibility that the structures with
$\tau$ slightly different from 1/4 may be able to develop a net moment
in the basal plane. The size of the effect is illustrated in Fig.\ 7,
which shows the calculated lock-in interval as a function of $\theta$
in a field of 30 kOe. $\Delta T$ increases like $\sqrt{\theta}$ at
small values of $\theta$ and is about 12 K at $\theta=30^\circ$.
Between 30$^\circ$ and 40$^\circ$ the helix is only stable in part of
the interval and is replaced by the fan structure in the other part.
For $\theta$ larger than about 40$^\circ$ only the fan is stable and
$\Delta T$ decreases to about 1 K at $90^\circ$. As mentioned above
the hexagonal anisotropy is very small at these temperatures and for
$\theta$ larger than about 0.5$^\circ$ the results shown in Fig.\ 7
are independent of whether the field is lying in the $a$--$c$ or
the $b$--$c$ plane. The transitions between the 1/4-structure and 
the surrounding incommensurable structures are established via the
creation of domain walls and are continuous, equivalently to
the case of $\tau$ close to 1/5.  

The vertical slope of $\Delta T$ as a function of $\theta$ at the
origin means that even the slightest deviation of the field from
perfect alignment along the $c$-axis will produce a sizable 
lock-in effect, e.g.\ $\Delta T$ is calculated to be 2.7 K at
$\theta=1^\circ$ when $H=30$ kOe. Both this value and the very weak
lock-in effect at zero field are in good agreement with the      
observations made by Noakes, Tindall and
collaborators,\cite{Noakes,Tin1} who saw no sign of a lock-in at zero
field but detected plateaus at $\tau=1/4$ with a temperature range of
2--2.5 K in a $c$-axis field of 17--30 kOe. The lock-in effect is
independent of the direction of the field component in the basal
plane, but it is not necessarily easy for the 1/4-structure to adjust
itself to even a slow spatial variation in the direction of the field, 
because a rotation of the ferromagnetic moment in the basal plane
requires a shift in the (average) phase angle for the 1/4-structure
which is about one quarter of the angle the moment is rotated. This 
effect, in combination with the presence of magnetic domains, may
explain why Venter {\it et al.}\cite{Venter1} did not detect any clear
plateau at $\tau=1/4$ in a $c$-axis field. 

The mean-field properties of the 36-layered, $\tau=5/18$, structure
just below the N\'eel temperature have also been
investigated. At zero field and at a temperature of 125 K, the model
predicts only a marginal lock-in effect, of the order of 0.05 K. The
calculations indicate an increase of the effect when a field is
applied in the basal plane, but the increase is not substantial, only
about a factor 1.5 in a field of 30 kOe. At this field it is assumed
that the structure is a fan, as the helix-fan transition is estimated
to occur at a field of about 20 kOe. The lock-in effects predicted at
these temperatures are far below what might be considered to
be observable effects, in contradiction with the experimental results.
Neutron diffraction experiments show a clear lock-in effect for the 
5/18-structure between 126 K and $T_{\text N}\simeq132.9$ K at a
field of 30 kOe along the $b$-direction,\cite{Tin4,Tin7} which
is accompanied by ultrasonic anomalies in the propagation of
longitudinal sound waves along the $c$-direction.\cite{Venter2}            
This lock-in phenomenon so close to the ordering temperature lies 
outside the range of what might be explained by a mean-field model. In
the mean-field approximation the anisotropy energies are nearly
eliminated close to $T_{\text N}$ due to thermal fluctuations. However,
it might be possible that these fluctuations behave somewhat
systematically such as to favor commensurable structures, analogously
to the commensurable effects induced by quantum fluctuations according
to the analysis of Harris {\it et al.}\cite{Harris}     

\section{Discussion and conclusion}
\label{sec:level4}
The most important result of the present investigation of the
commensurable structures in holmium is that the increased stability
of the 10-layered periodic structure around 42 K and of the 8-layered 
periodic structure around 96 K, observed when applying a field along the
$c$-axis, can be understood. In both cases the explanation relies
totally on the trigonal coupling, adding to the evidence for the
presence of this coupling in holmium. The two cases are also the
only ones found where the trigonal coupling has any
significant effect on the stability of the commensurable structures.
  
With a few exceptions the calculated ranges in which the different
commensurable structures are stable, are larger than indicated by the
experiments, and these differences are most noticeable at 
low temperatures. This may be explained by the occurrence of metastable
structures in the samples. The neutron diffraction experiments of 
Cowley{\it et al.}\cite{Cow2} show that the crystals may contain
several domains with different structures below 40 K. As the
temperature is raised the energy barriers between the metastable
structures decrease and the thermal energies increase, so that the
system may more easily reach thermal equilibrium. At low temperatures,
in the regime of the spin-slip structures, there is the additional
possibility that the regularly spaced spin-slip layers in
the equilibrium state are disordered to some extent. The x-ray
diffraction measurements of Helgesen {\it et al.}\cite{helg} 
indicate that this is the case. They have observed a reduction of the
longitudinal correlation length between 40 and 20 K by a factor of 
three, a reduction which is partly removed when the spin-slip layers
disappear at the lock-in transition to $\tau=1/6$ at about 20 K. 

The only indisputable discrepancy between the theory and the
experiments is found in the behavior displayed by the
5/18-structure. The experiments\cite{Tin4,Tin7,Venter2} indicate that 
this structure locks-in between 126 K and $T_{\text N}$ in a $b$-axis
field of 30 kOe, whereas the calculations only show a marginal effect.
This discrepancy does not necessary question the model, but is more likely
a consequence of the limited validity of the mean-field approximation 
close to the magnetic phase transition. The fluctuations neglected in
the calculations may be so large in this case that their contribution
to the free energy is decisive for the commensurable effects.   

At temperatures well below $T_{\text N}$ we do not expect any
major corrections to the lock-in intervals derived in the 
mean-field approximation. Any discrepancy would rather be due to a
failure of the model than to the use of this approximation. The model
does not include magnetoelastic effects, except implicitly through the 
variation of ${\cal J}(q)$. The temperature dependence of the exchange
interaction, which is accounted for in a phenomenological way
in consistency with the behavior of the spin-wave energies, may
relate to changes of the $c/a$-ratio.\cite{And} The magnetoelastic 
effects in holmium, like in the other rare-earth metals, are
important, but they do not seem to have much direct influence on the
lock-in phenomena. Possible exceptions are the cases of the
$\tau=2/11$ and 3/16 spin-slip structures, which both have the
distinct property that the averaged value of the quadrupole moments in
the basal plane is non-zero. The neglect of magnetoelastic effects
in the model implies that all the parameters are assumed to stay
constant as functions of the field, at a fixed temperature. A field
dependence of ${\cal J}({\text \bf q})$ is expected based on the
Elliott--Wedgwood theory,\cite{Elliott} because the polarization of
the conduction electrons is changed by the field. It is therefore
interesting to notice that the model is able to account for most of 
the field dependence of the ordering wave vector in holmium, observed
both at low temperatures and at temperatures around 80--100 K, without
much need for invoking a field-dependence of the  exchange coupling.

The trigonal coupling derived in the present analysis, is somewhat
smaller than the coupling considered by Simpson {\it et
  al.},\cite{Simp} and in comparison with the exchange coupling its
relative magnitude is ten times smaller in holmium than in
erbium.\cite{Cow3} Nevertheless, the two commensurable effects in
holmium determined by the trigonal coupling are found to be very
pronounced, and a more detailed experimental investigations of these
two effects would be valuable. The lock-in temperature interval of the 
1/5-structure is predicted to be larger than indicated by the
variation in the position of the first harmonic,\cite{Noakes,Tin1} and
a study of the behavior of the fifth or seventh harmonics will be
useful for a clarification of the experimental situation. The strong
lock-in of the 1/4-structure around 96 K indicated by the mean-field
model, Fig.\ 7, deserves further studies, in which the field is
applied by purpose in a direction making a non-zero angle with the
$c$-axis, or with the basal-plane. 
 
 \begin{acknowledgments}
The author wants to express his gratitude to Allan R. Mackintosh who
tragically died after a car accident. He was always extremely helpful
and full of insight, and the author and many others have benefited from
discussions with him on problems in physics. We will all remember and
miss his enthusiasm and deep understanding of the physics of metals
and of the rare earths.    
\end{acknowledgments}

\begin{table}
\caption{The trigonal coupling parameters ($10^{-4}$meV).}
\label{table2}
\begin{tabular}{cccc}
$n$ & 1 & 2 & 3 \\
\noalign{\vspace{1pt}}
\tableline
\noalign{\vspace{2pt}}
$\big[K{}_{31}^{21}\big]_n^{}$ & 0.7 & 0.4 & 0.2 \\
\end{tabular}
\end{table}
\begin{table}
\caption{The crystal-field parameters (meV).}
\label{table1}
\begin{tabular}{cccc}
$B_2^0$ & $B_4^0$ & $B_6^0$ & $B_6^6$ \\
\noalign{\vspace{1pt}}
\tableline
\noalign{\vspace{2pt}}
0.024 & 0.0 & $-0.95$ $10^{-6}$ & 9.4 $10^{-6}$ \\
\end{tabular}
\end{table}
\begin{table}
\caption{The inter-planar exchange parameters (meV) as functions of
  temperature.}
\label{table3}
\begin{tabular}{rccccccc}
$T$ [K] & ${\cal J}_0$ & ${\cal J}_1$ & ${\cal J}_2$ & ${\cal J}_3$ 
& ${\cal J}_4$ & ${\cal J}_5$ & ${\cal J}_6$ \\
\noalign{\vspace{1pt}}
\tableline
\noalign{\vspace{2pt}}
0 &0.300&0.09&0.006&$-0.0140$&$-0.006$&$-0.002$&$\phantom{-}0.0\phantom{00}$\\
50&0.290&0.10&0.010&$-0.0290$&$-0.005$&$\phantom{-}0.008$&$-0.004$\\
72&0.267&0.11&0.010&$-0.0377$&$-0.001$&$\phantom{-}0.004$&$-0.003$\\
96&0.245&0.11&0.010&$-0.0463$&$\phantom{-}0.006$&
$\phantom{-}0.0\phantom{00}$&$\phantom{-}0.0\phantom{00}$ \\
125&0.210&0.11&0.010&$-0.0640$&$\phantom{-}0.006$&
$\phantom{-}0.0\phantom{00}$&$\phantom{-}0.0\phantom{00}$ \\
\end{tabular}
\end{table}

\begin{figure}
\caption{The ordering wave vector in Ho as a function of temperature
below 50 K. The calculated results are shown by the horizontal solid 
lines, which are connected with vertical thin solid or thin dashed 
lines corresponding respectively to the results obtained at zero or at
a field of 10 kOe applied along the $c$-axis. The symbols show the
experimental results of Cowley {\it et al.}$^{27}$ at various values
of the $c$-axis field as defined in the figure. The thick dashed line,
between 35 and 48 K, indicates the variation of $\tau$ derived by
Tindall {\it et al.}$^{22}$ from the position of the primary magnetic
diffraction peak in a $c$-axis magnetic field of 30 kOe. The smooth
curve shown by the thin solid line is the temperature dependent
position of the maximum in ${\cal J}(q)$, as determined by Eq.\ (3.1).}   
\label{fig1}
\end{figure}
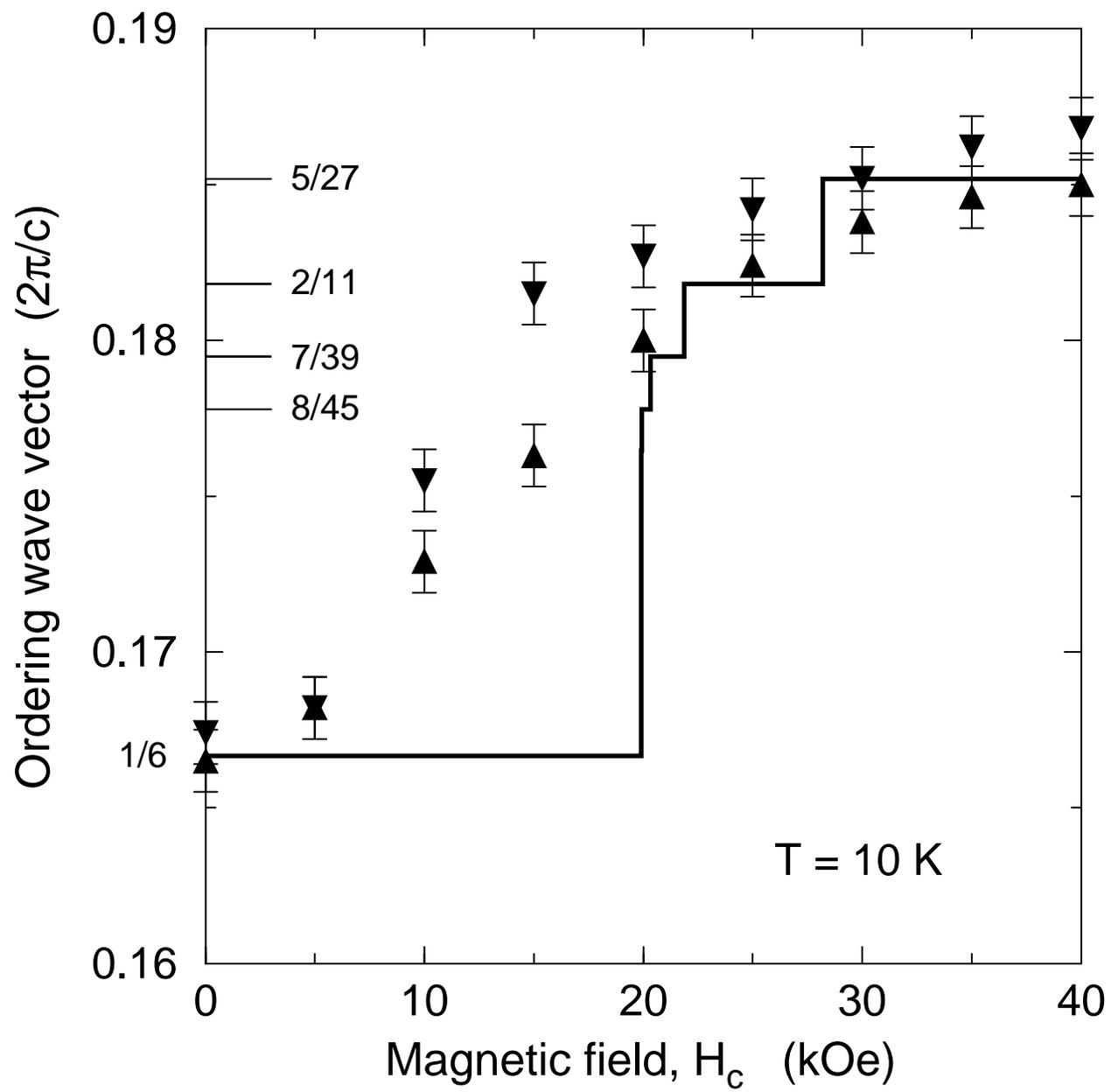
\begin{figure}
\caption{The ordering wave vector in Ho as a function of a $c$-axis
field at 10 K. The solid line is the theoretical result, and the
triangles pointing up or down are the experimental results of Cowley
{\it et al.}$^{27}$ obtained at increasing or decreasing field values,
respectively. The systematic difference between the two sets of
results indicates that hysteresis effects are important. The
experimental results are derived from the position of the principal
magnetic satellite, and thus include implicitly an averaging of the
ordering wave vector over the different domains.}  
\label{fig2}
\end{figure}
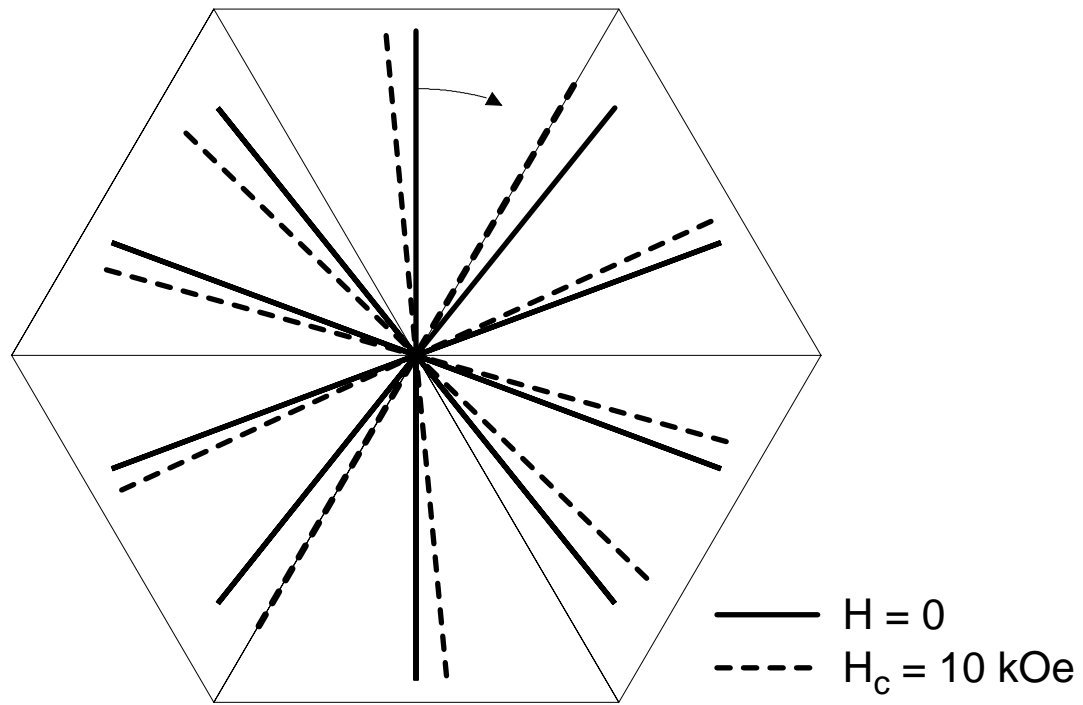
\begin{figure}
\caption{The hodograph of the basal-plane moments of the 10-layered
(221)-structure calculated at 42.185 K in zero field and in the
presence of a field of 10 kOe along the $c$-axis. The two hodographs are
rotated about 30$^\circ$ relatively to each other as indicated by the
arrow.}
\label{fig3}
\end{figure}
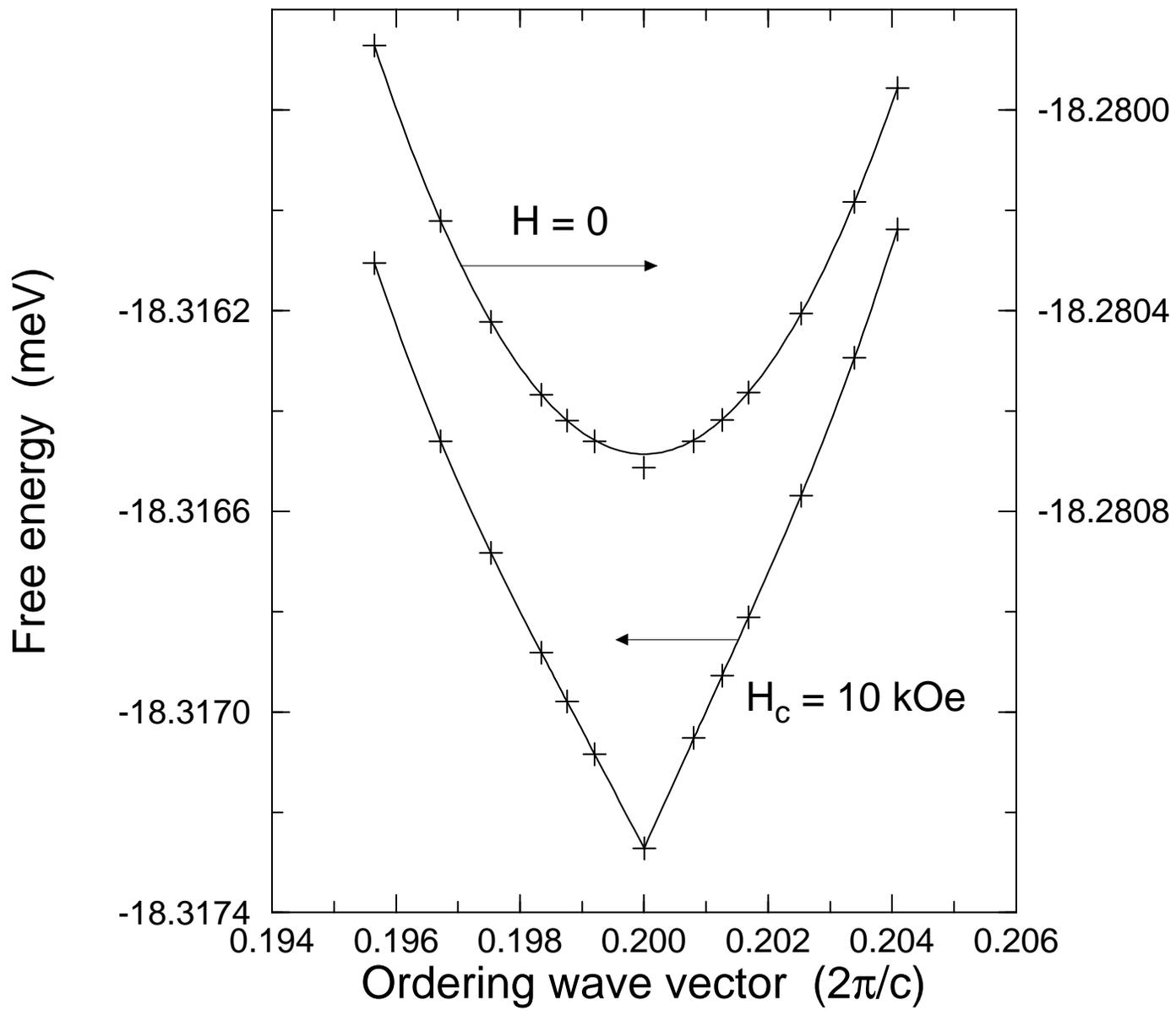
\begin{figure}
\caption{The free energy of the commensurable structures at 42.185 K
as a function of the ordering wave vector calculated at zero field and
at a field of 10 kOe applied along the $c$-axis. The energy scales used
in the two cases are displaced with respect to each other as
indicated on the figure. The plus signs denote the free energy of the
structures calculated with commensurable periods between about 100 and 500
hexagonal layers.} 
\label{fig4}
\end{figure}
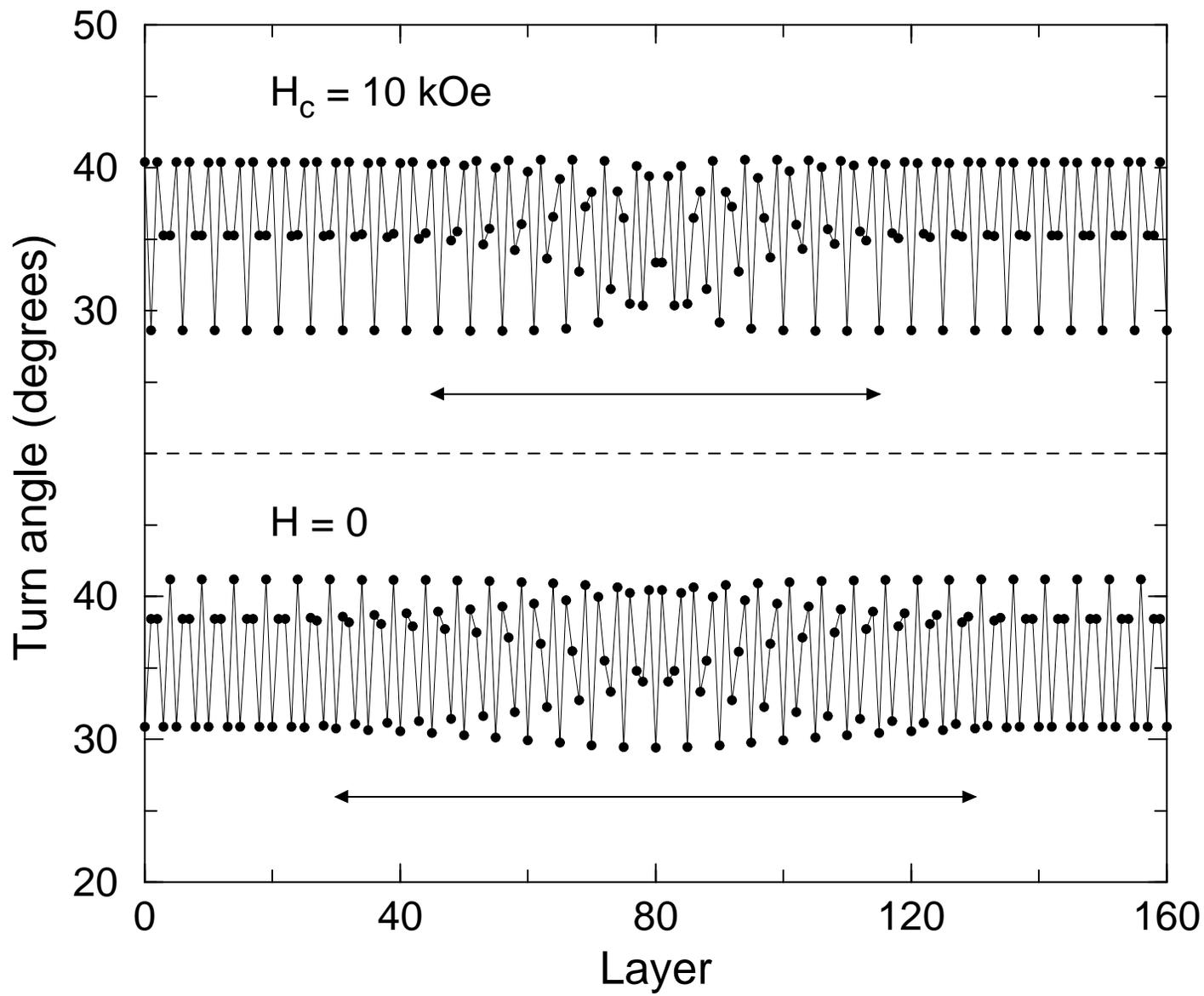
\begin{figure}
\caption{The turn angles of the moments in different hexagonal layers
  numbered along the $c$-axis, calculated at zero field and in the 
present of a $c$-axis field of 10 kOe at the same temperature, 42.185 K,
as considered in the Figs.\ 3 and 4. The domain walls indicated
by the arrows lead to negative phase shifts of the regular 1/5-structural
parts on each side of the walls, which are $12^\circ$ at zero field 
and $24^\circ$ in the case of $H_c=10$ kOe.} 
\label{fig5}
\end{figure}
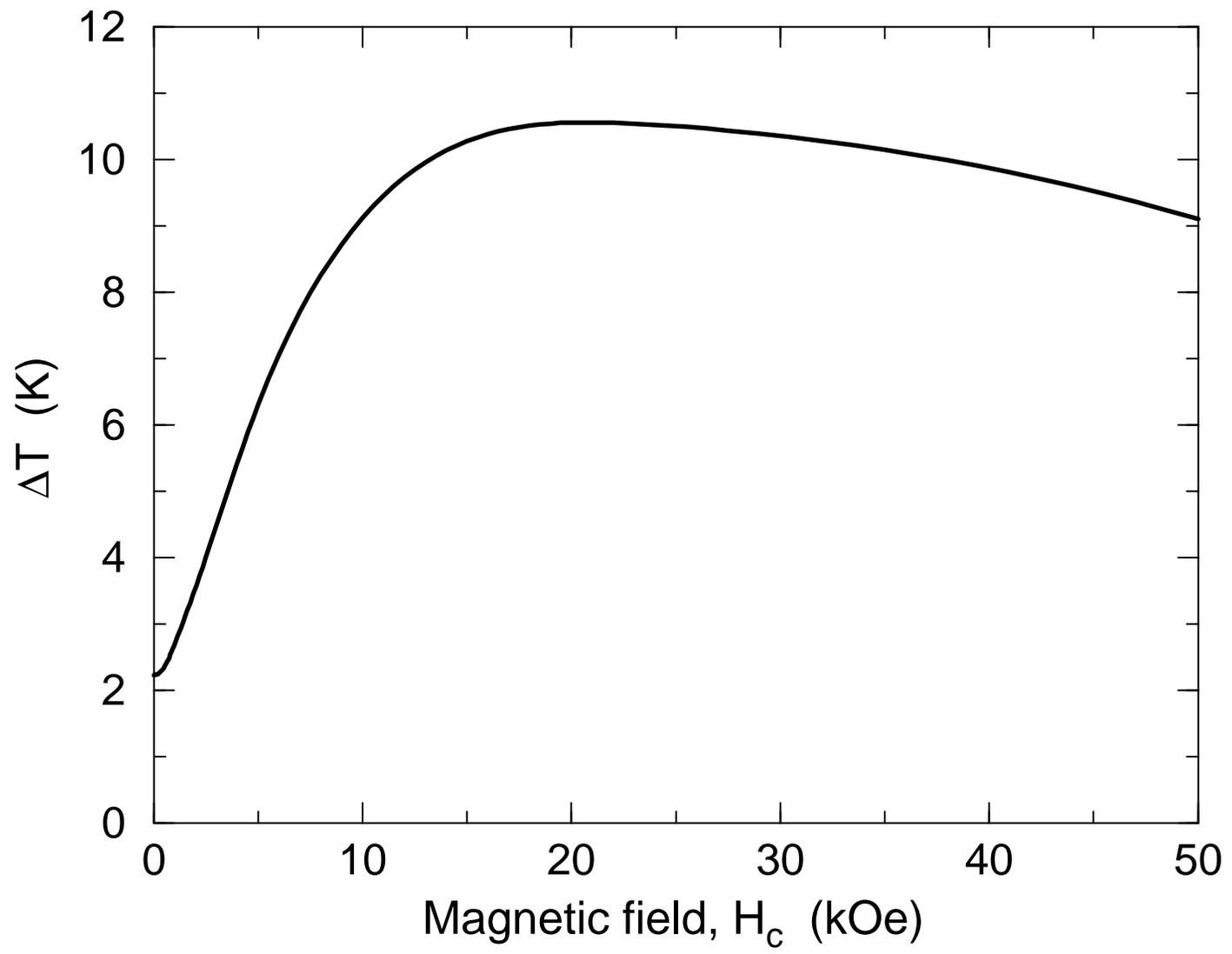
\begin{figure}
\caption{The calculated lock-in temperature interval, $\Delta T$, of the
  1/5-phase as a function of the $c$-axis field.} 
\label{fig6}
\end{figure}
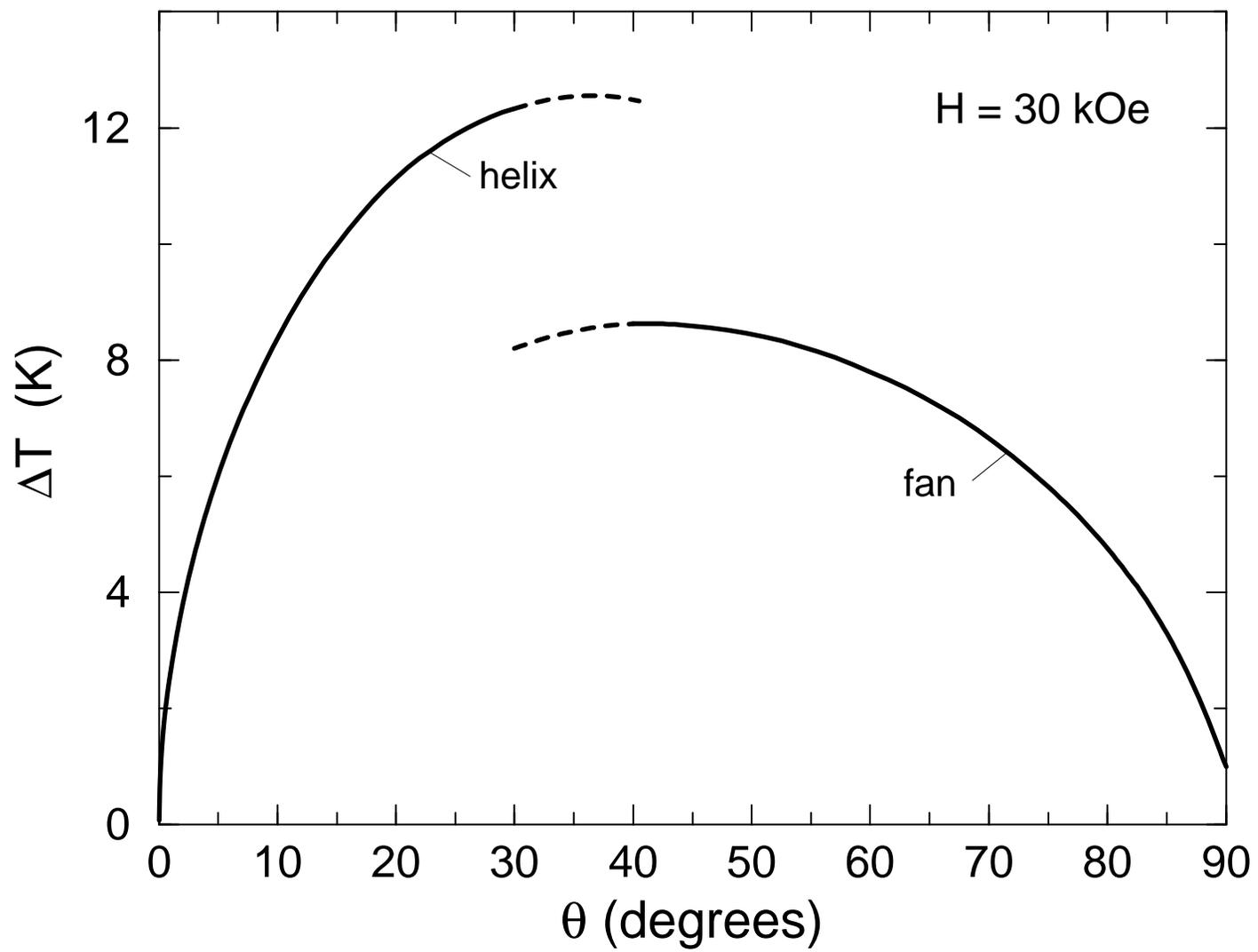
\begin{figure}
\caption{The calculated lock-in temperature interval, $\Delta T$, of the
1/4-phase as a function of a field of 30 kOe applied in a direction
making the angle $\theta$ with the $c$-direction. The midpoint
of the interval is between 95--97 K in the helical case and lies
between 102--100 K in the fan phase.} 
\label{fig7}
\end{figure}

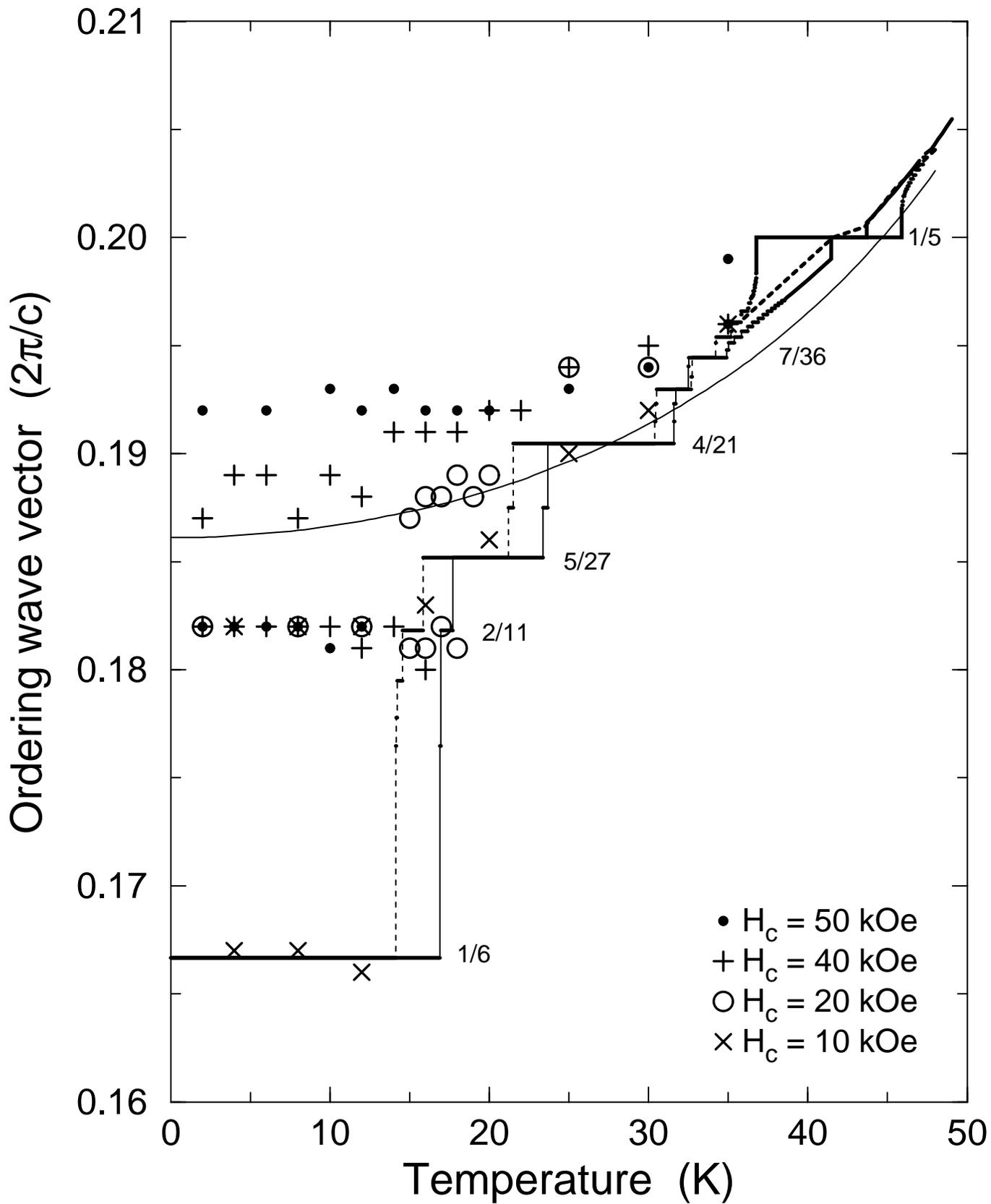
\end{document}